\documentclass[11pt]{article}
\usepackage{moriond,epsfig}

\def\Journal#1#2#3#4{{#1} {\bf #2}, #3 (#4)}

\def\NIM{\em Nucl. Instrum. Methods}

\def\NPB{{\em Nucl. Phys.} B}
\def\PLB{{\em Phys. Lett.}  B}
\def\PRL{\em Phys. Rev. Lett.}
\def\PRD{{\em Phys. Rev.} D}

\def\be{\begin{equation}}
\def\ee{\end{equation}}
\def\bea{\begin{eqnarray}}
\def\eea{\end{eqnarray}}

\begin{document}
\vspace*{4.cm}
\title{HEAVY FLAVOR PHYSICS : LIFETIMES AND FLAVOR CHANGING NEUTRAL CURRENTS}

\author{ S. MALDE, on behalf of the CDF and DO collaborations }

\address{Department of Physics, DWB, Keble Road, Oxford OX1 3RH, UK}

\maketitle
\abstracts{The Tevatron, with $p\overline{p}$ collisions at $\sqrt{s}=$1.96 TeV, can produce all flavors of $B$ hadrons and allows for unprecedented studies in the $B$ physics sector. These range from measurements of $B$ hadron properties to searches of new physics signatures. The CDF~\cite{cdfdetector} and D0~\cite{d0detector} detectors currently have more than 7 fb$^{-1}$ of data recorded. This paper presents a selection of recent results on lifetimes and flavor changing neutral currents using between 4.3--5.0 fb$^{-1}$ of data.}

\section{$B$ hadron lifetimes}

The experimental measurement of $B$ hadron lifetime ratios is an important test
of the theoretical approach to $B$ hadron observables known as the heavy quark expansion. The ratio $\tau(B^+)/\tau(B^0)$ (charge conjugates are implied throughout) is predicted~\cite{PDG}$^,$\cite{Bigi}$^,$\cite{NLOQCD}$^,$\cite{Subleading} to be in the range 1.04--1.08 and the ratio $\tau(\Lambda^0_b)/\tau(B^0)$ in the range 0.83--0.95.\cite{PDG}$^,$\cite{Subleading}$^,$\cite{LBOriginals} The measured world average $B^+$ and $B^0$ lifetimes are dominated by the Belle experiment.\cite{ref:BelleResult} Of recent interest is the $\Lambda^0_b$ lifetime. Until 2006 all measurements were in agreement but lay at the lower end of the theoretically expected value. Since then, two high precision CDF measurements are significantly above previous results.\cite{ref:Mark}$^,$\cite{ref:Petar}  The analysis described here is the most precise measurement of the $B^+$, $B^0$, and $\Lambda^0_b$ lifetimes and ratios. 

\begin{figure}
\begin{center}
\psfig{figure=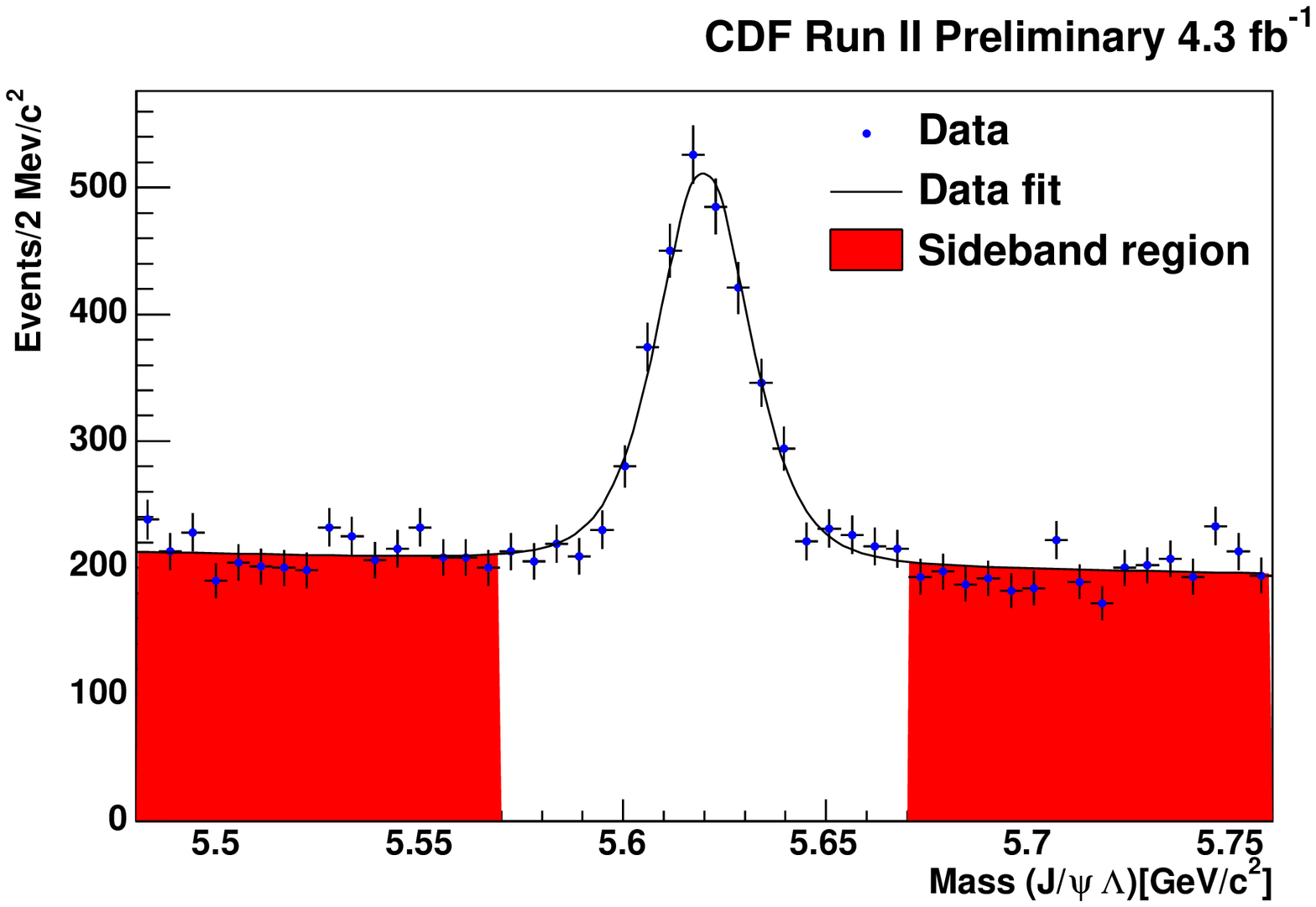,height=1.8in}
\psfig{figure=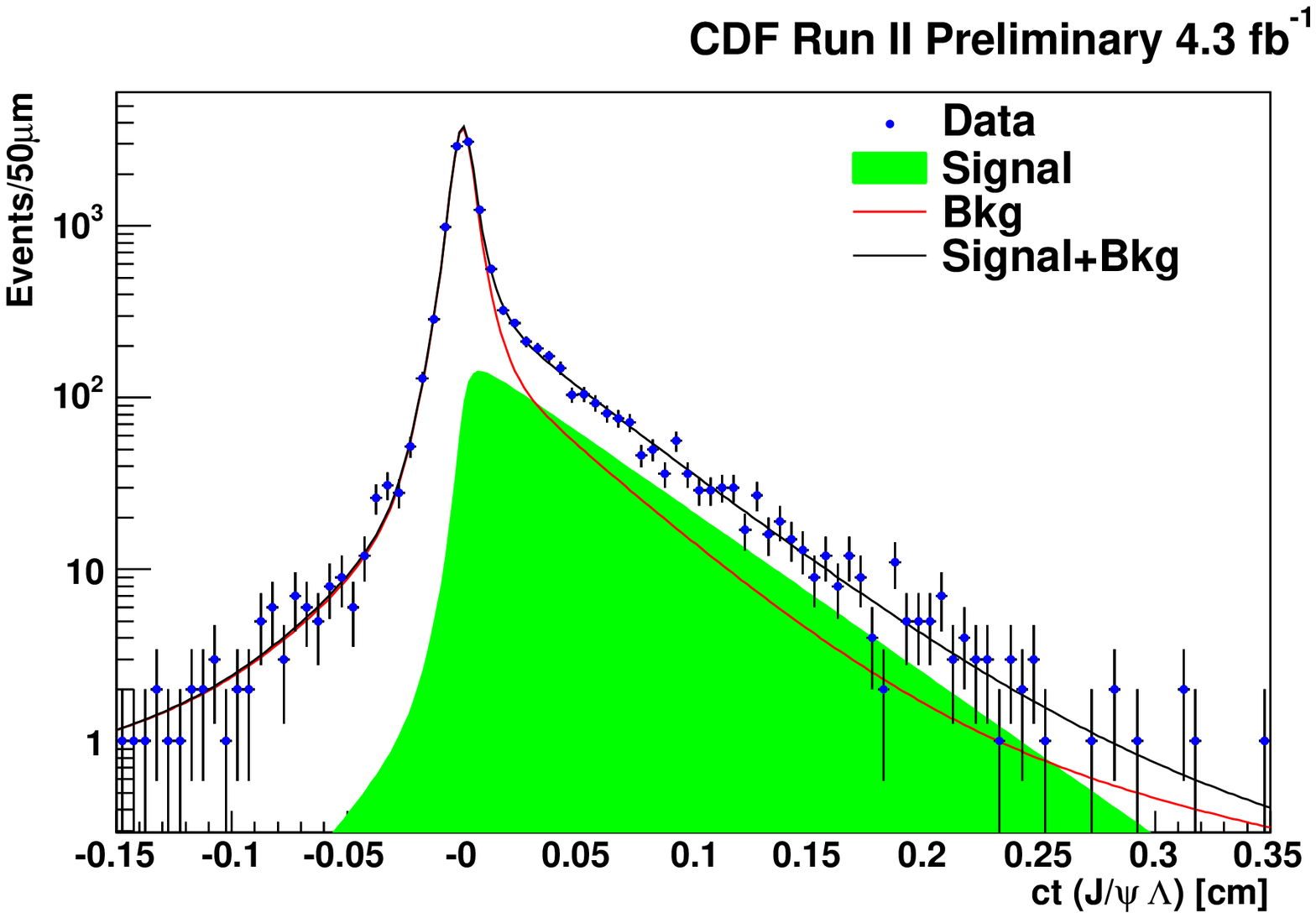,height=1.8in}
\caption{The $\Lambda_b$ mass and proper decay time distributions.
\label{blifes}}
\end{center}
\end{figure}

The $B^+$, $B^0$, and $\Lambda_b$ lifetimes were measured using 4.3 fb$^{-1}$ of data with decay channels $B^+ \to J/\psi K^+$, $B^0 \to J/\psi K^*$, $B^0 \to J/\psi K_{s}$ and $\Lambda_b \to J/\psi \Lambda$. In previous measurements the uncertainty due to detector resolution has been a leading source of systematic uncertainty. In this analysis the proper decay time is determined using the $J/\psi$ vertex to provide similarity in the decay time resolution between channels and to allow for the cancellation of certain systematic uncertainties. A detailed resolution model is also introduced in this analysis. The signal decay time is modelled as an exponential decay convolved with the resolution model.  The resolution model is a superposition of three Gaussians. They are each centred at $t=0$, and have a width of event decay time uncertainty, $\sigma^{ct}_i$, multiplied by a scale factor. The restriction to models symmetric about $t=0$ is motivated by simulation, while the number of components is determined from data. The parameters of the resolution function are determined from the mass sidebands as the fraction of background events expected to originate from the primary vertex is between 80-90$\%$, depending on channel and background model, and therefore provides a useful sample from which to determine the resolution. The overall fit is an unbinned likelihood fit to the mass, decay time and decay time uncertainty distributions simultaneously. The projections of the mass and decay time distributions from the $\Lambda_b$ data are shown in Fig.~\ref{blifes}. 

We measure $\tau_{B^+} = 1.639 \pm 0.009 ~({\rm stat}) \pm 0.009~{\rm (syst)~ ps}$, $\tau_{B^0} = 1.507 \pm 0.010 ~({\rm stat}) \pm 0.008~{\rm (syst)~ ps}$, and $\tau_{\Lambda^0_b} = 1.537 \pm 0.045 ~({\rm stat}) \pm 0.014~{\rm (syst)~ ps}$. The lifetime ratios are calculated as  $\tau_{B^+}/\tau_{B^0} = 1.088  \pm 0.009~({\rm stat})\pm 0.004~({\rm syst})$ and $\tau_{\Lambda^0_b}/\tau_{B^0}= 1.020   \pm 0.030~({\rm stat})\pm 0.008~({\rm syst}).$~\cite{lifepub} These are the world's best measurements of the lifetimes and ratios. The improvement in the systematic uncertainty from 0.033 ps (1.0 fb$^{-1}$ to 0.014 ps (4.3 fb$^{-1}$) is evident in the $\tau(\Lambda_b)$ measurement. The $\Lambda_b$ lifetime remains higher than the world average but is not inconsistent with theoretical predictions.  

Measurements of other $B$ baryon lifetimes are also interesting as they can validate the prediction of the baryon hierarchy and ratios. CDF and D0 have both observed the $\Xi_b$\cite{chib}$^,$\cite{chib2} and the $\Omega_b$\cite{omegad0}$^,$\cite{omegacdf} baryons. CDF has performed lifetime measurements on these samples. The low statistics, particularly in background motivates a different approach to lifetime measurement. The data are divided into bins of proper decay time. A mass fit in each bin determines the number of signal candidates in each bin. The lifetime is then determined from the yield in each bin compared to the expected distribution for a given lifetime. The method has been validated on $\Lambda_b$ and $B^0$ decay modes and gives results consistent with those from other methods. Using 4.3 fb$^{-1}$ CDF measures $\tau (\Xi_b) = 1.56^{+0.27}_{-0.25} (\mathrm{stat}) \pm 0.02(\mathrm{syst})$ ps which is the first measurement using a fully reconstructed decay of $\Xi_b$, and $\tau(\Omega_b)=1.13^{+0.53}_{-0.40}(\mathrm{stat}) \pm 0.02 (\mathrm{syst})$ ps which is the first measurement of $\tau(\Omega_b$)\cite{omegacdf}.

\section{Flavor Changing Neutral Currents}

Flavor changing neutral currents (FCNC) are forbidden at tree-level in the standard model. Processes that involve FCNC through higher order diagrams can provide information on new physics. New particles participating in the higher order diagrams can enhance decay rates or alter expected kinematic distributions from the standard model predictions. FCNC provide a complementary approach to new physics searches alongside direct searches, as any signatures of new physics that are observed give information on the flavour structure of new physics, and thus constrain the set of new physics models that are consistent with data observations.

\subsection{$B \to \mu^+ \mu^-$}

One highly studied decay is $B_s \to \mu^+ \mu^-$. The standard model branching fraction is\cite{bmumutheory} (3.42 $\pm$ 0.54)$\times 10^{-9}$ which is beyond the CDF and D0 detector experimental sensitivity.However, these branching ratios can be enhanced by a factor of $10-100$ by supersymmetric models or other new physics. The decay $B^0 \to \mu^+ \mu^-$ is further suppressed by the ratio of CKM elements, $|V_{td}/V_{ts}|^2$. Given the detector sensitivities, observation of either decay is an unequivocal signature of new physics. If observed, the ratio of measured $B_s$ and $B^0$ branching fractions would give information on the flavor structure of the new physics.
 
\begin{figure}
\begin{center}
\psfig{figure=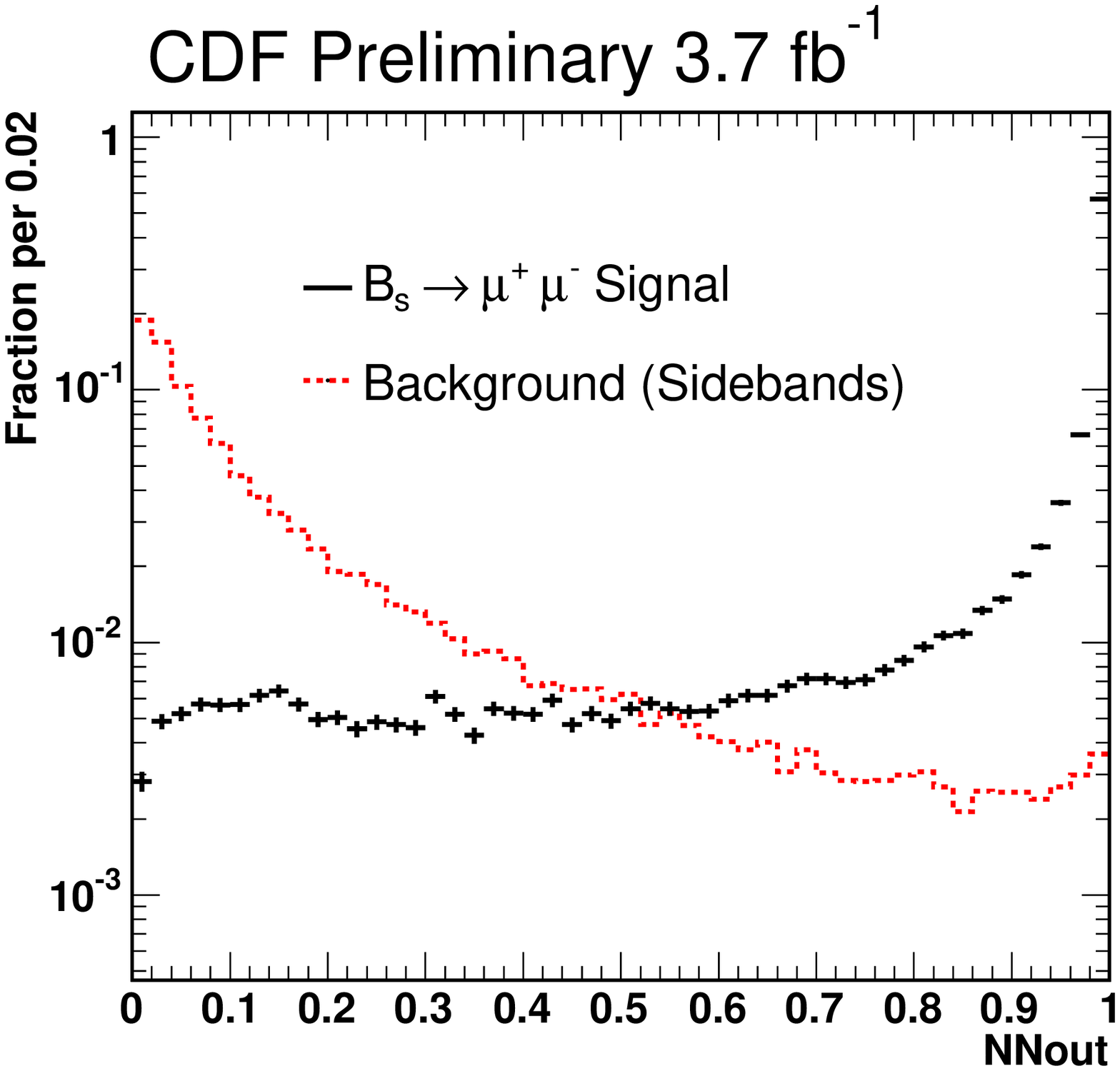,height=2.0in}
\psfig{figure=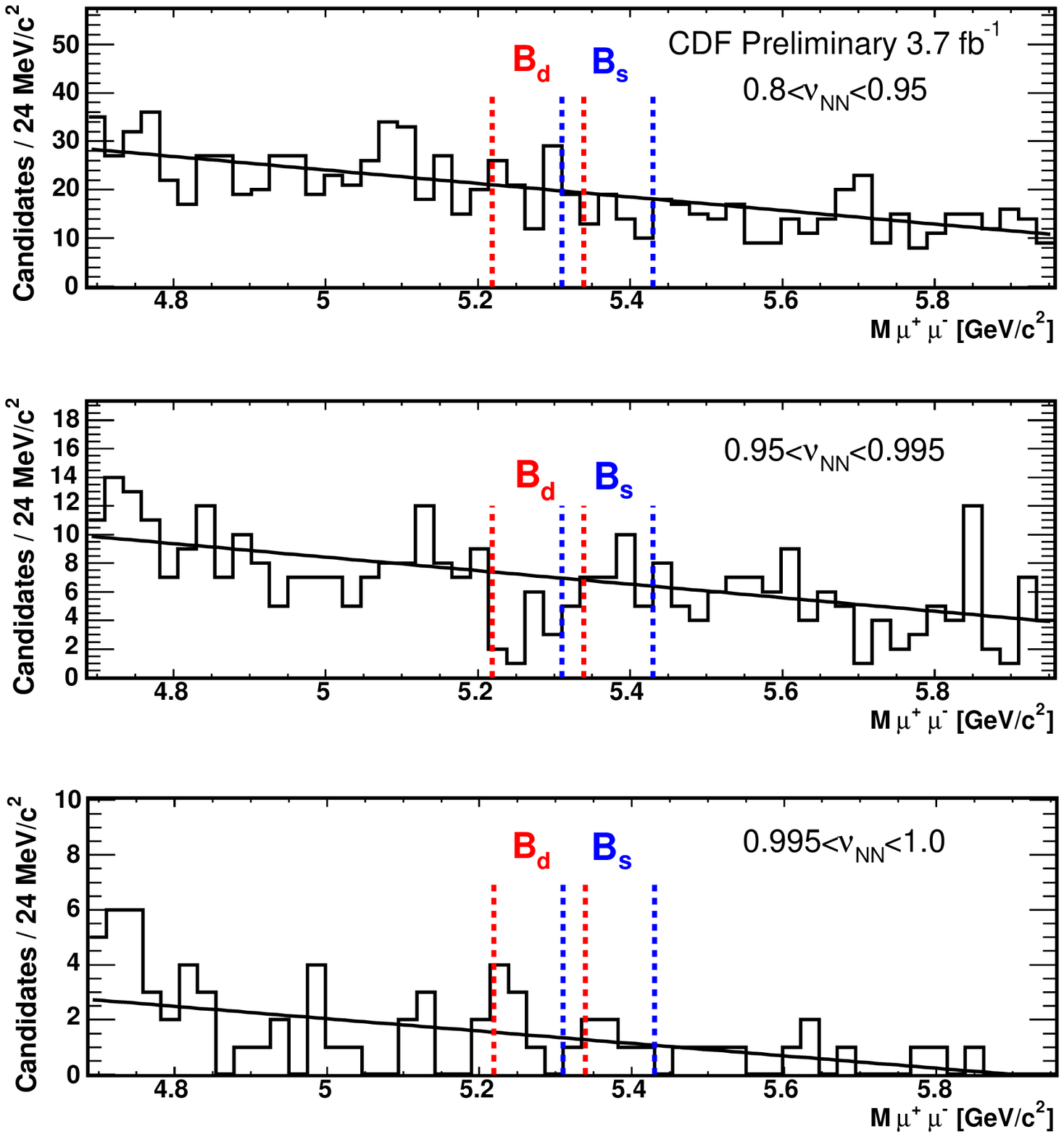,height=2.0in}
\caption{The left plot shows the neural network discriminant for signal and background. The right plot shows the data after selection in three bins of the neural network discriminant.
\label{nnout}}
\end{center}
\end{figure}

The analyses carried out at CDF and D0 are similar; the CDF analysis uses 3.7 fb$^{-1}$ and is described below. One challenge in this analysis is to reduce the large backgrounds. A number of baseline selection requirements are applied which result in a reduction of background by a factor of 300 while 50$\%$ of signal would remain in the geometric and kinematic acceptance of the detector. The baseline selection includes selection on transverse momentum, vertex quality, and  muon ID algorithms which reduce backgrounds where hadrons have been misidentified as muons and decays involving a kaon that has decayed inflight to a muon. For further enhancement of signal events a neural network is used. It is trained on six variables; the proper decay time and proper decay time significance, the transverse momentum of the di-muon candidate, the B-candidate track isolation, the $p_T$ of the lower momentum muon candidate, and the 3D opening angle between the vectors $\vec{p}^{\mu\mu}$ and the displacement vector between the primary vertex and the dimuon vertex. The neural network is trained using signal events generated by Monte Carlo and mass sideband events for background. The neural network output is shown in Fig.~\ref{nnout}. The remaining background is estimated from continuum combinatorics from sidebands and mis-reconstructed $B \to h h $ decays which peaks in the signal region. The background estimates are cross-checked using control samples from data such as like-sign muons. The background predictions are compared with the observed data and no statistically significant discrepancies are observed.

A relative normalisation to the channel $B^+ \to J/\psi K^+$ is used to determine the $B \to \mu \mu $ branching fraction. The number of observed signal events, $N_s$, and normalization events, $N_+$, can be used to obtain the branching fraction via:
\begin{equation}
\mathcal{B}(B^0_s \to \mu^+ \mu^-) = \frac{N_s}{N_+}\cdot\frac{\epsilon_+}{\epsilon_s}\cdot\frac{f_u}{f_s}\cdot \mathcal{B}(B^+ \to J/\psi K^+, J/\psi \to \mu^+ \mu^-), 
\end{equation}
where $f_u$, $f_s$ are fragmentation fractions and $\epsilon_+$ and $\epsilon_s$ are the efficiencies for the normalisation and signal channels. The efficiencies are determined through a combination of Monte Carlo simulation and data driven techniques involving samples of $J/\psi$, $B \to J/\psi K^+$ and $B \to J/\psi \phi$.

After selection, the data observed in the mass ranges corresponding to $B_s$ and $B^0$ are consistent with there being only background events and are shown in Fig~\ref{nnout} in three separate bins of the neural network discriminant. CDF sets limits on the branching fractions for these decay processes. Using a data sample of 3.7 fb$^{-1}$ CDF extracts a 95$\%$ (90$\%$) C.L for $\mathcal{B} (B_s \to \mu^+ \mu^-)<4.3 \times 10^{-8} (3.6) \times 10 ^{-8}$ and $\mathcal{B} (B^0 \to \mu^+ \mu^-)<7.6 \times 10^{-9} (6.0) \times 10 ^{-9}$.~\cite{cdfbmumu} These are the world's best limits on these branching ratios.
At D0 an analysis of 2 fb$^{-1}$ obtained a limit of  $\mathcal{B} (B_s \to \mu^+ \mu^-)<9.5 \times 10^{-8} (7.5) \times 10 ^{-8}$.~\cite{d0bmumu2} The analysis of 5 fb$^{-1}$ is on-going with the signal region still blinded. The expected upper limit on the branching fraction on these data is  $\mathcal{B} (B_s \to \mu^+ \mu^-)<5.3 \times 10^{-8} (4.3) \times 10 ^{-8}$.~\cite{d0bmumu5} These measurements at the Tevatron have limited the parameter space of allowed new physics models as large enhancements of the $B \to \mu \mu$ branching ratio are inconsistent with the observed data.
\subsection{$B \to \mu^+ \mu^- h$}
One group of interesting decays to study are $B \to \mu \mu h$ where the $B$ is either $B^+$, $B^0$ or $B_s$ and $h$ stands for either $K^+$, $K^*$ or $\phi$, respectively. The branching ratios of these decays are $\mathcal{O} (10 ^{-6})$ and are observable. New physics processes can enhance decay amplitudes. The interference between amplitudes due to new physics particles and the standard model processes may be observable in effect on branching ratios, polarisation and forward backward asymmetry. Comparison of these observables to standard model expectation can indicate whether the underlying dynamics are governed by the standard model or other models such as SUSY, or 4th generation quarks.

The three decay channels are reconstructed from data collected by triggers that require two charged particles with $p_T \geq 1.5$GeV/$c$ or 2.0 GeV/$c$ with corresponding hits in the muon chambers. Vertices comprising of the muon pair and the $h$ are required to have a $\chi^2$ probability of at least 10$^{-3}$. For the normalisation of the branching ratios the control channels $B \to J/\psi h$ are also reconstructed. The same final state allows a number of systematic uncertainties to cancel. Selection criteria are placed on the decay time significance, transverse momentum, impact parameter of the $B$ candidate, the $\phi$ and $K^*$ mass, and particle identification to reduce the combinatoric and fake muon backgrounds. The normalisation channels are required to have the di-muon invariant mass within 50 MeV/$c^2$ of the $J/\psi$ mass. For the signal channels, candidates with a di-muon mass in the region near the $J/\psi$ and $\psi'$ are rejected. Other criteria are applied to reduce other peaking backgrounds. After the loose selection a neural network technique is used to provide the final selection. It is optimised to maximise both the branching ratio and the lepton forward-backward asymmetry, $A_{FB}$, significance. 

\begin{figure}
\begin{center}
\psfig{figure=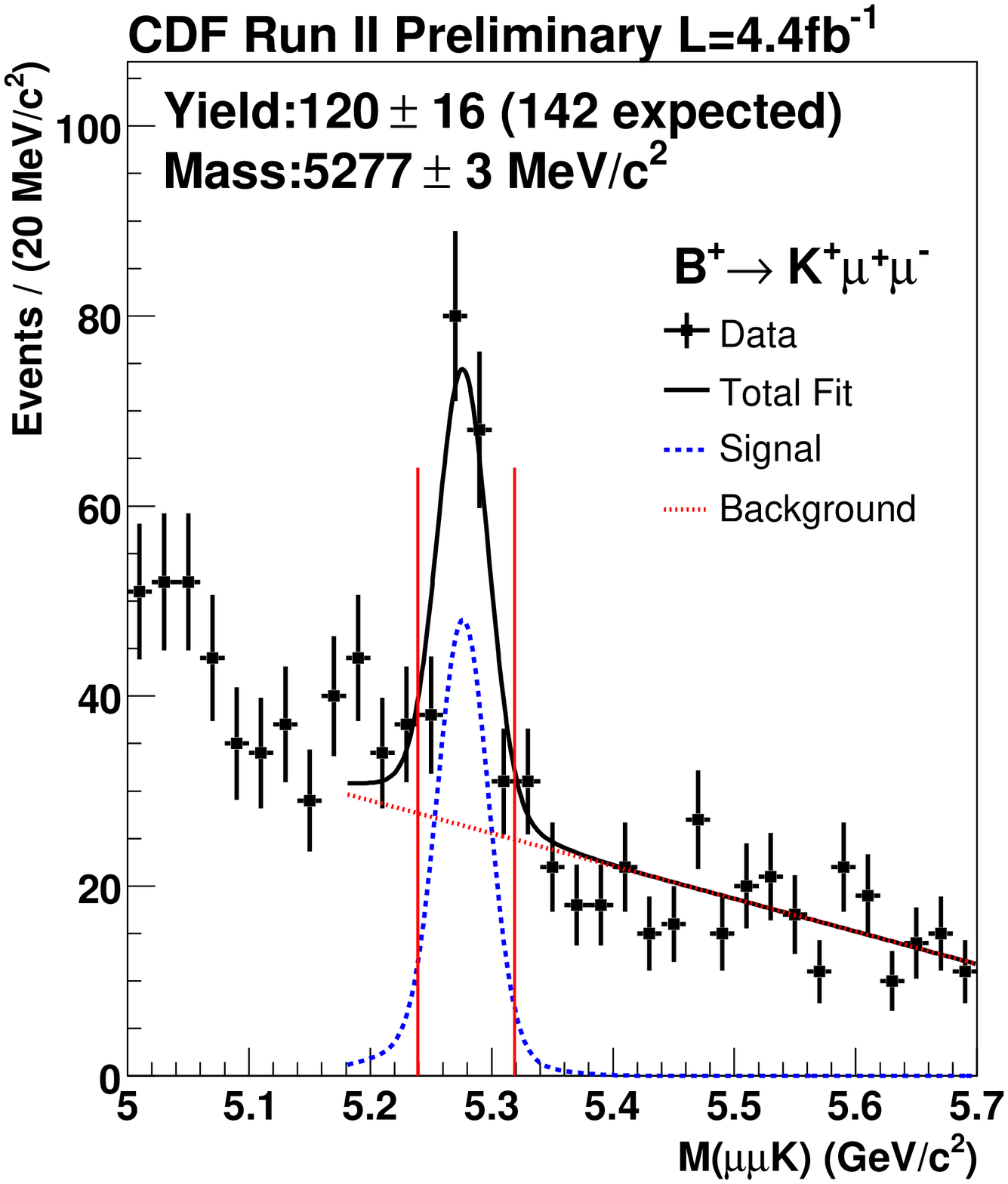,height=2.2in}
\psfig{figure=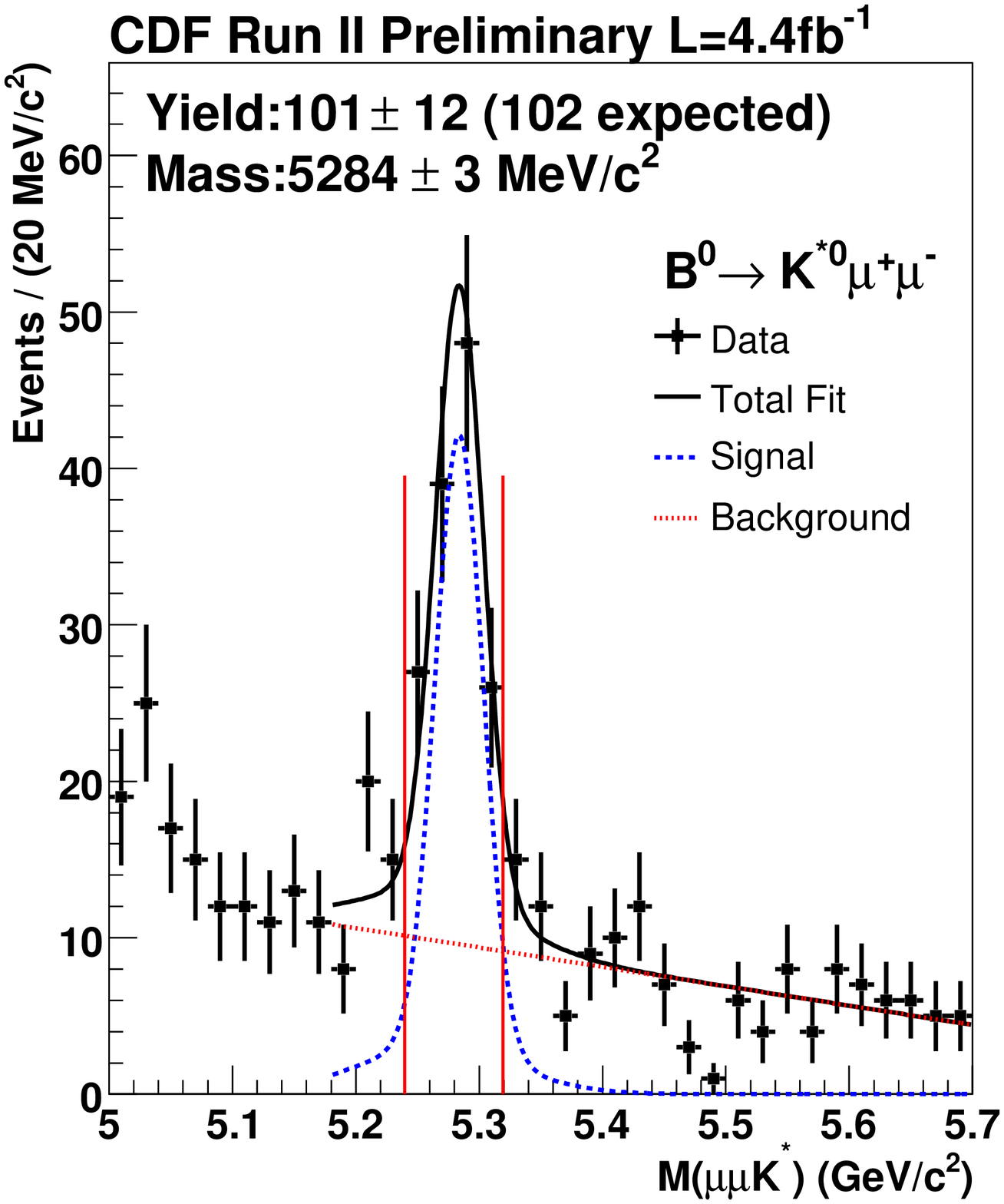,height=2.2in}
\psfig{figure=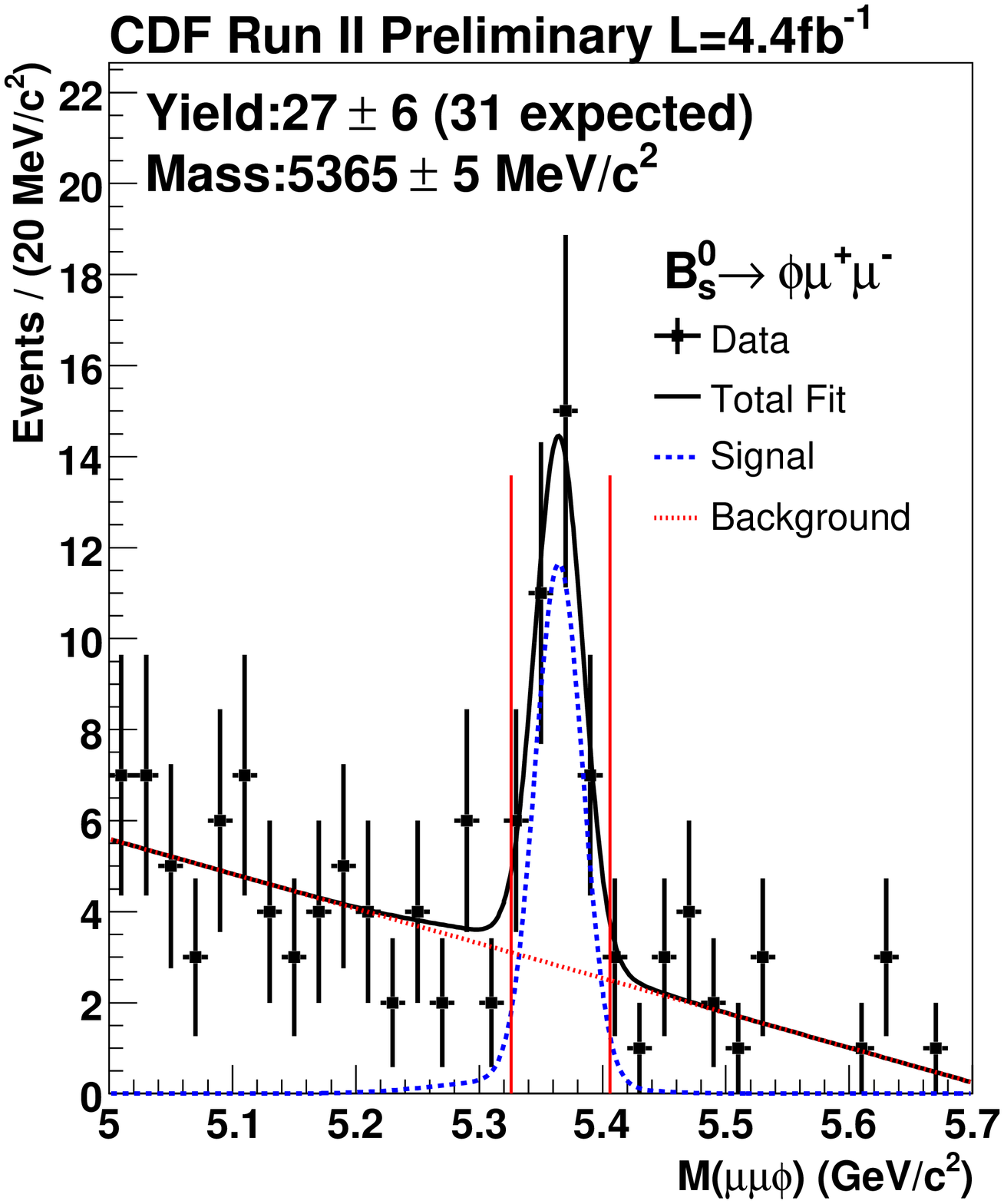,height=2.2in}
\caption{The mass distribution of $B \to \mu \mu h$ candidates after selection.
\label{bmumuhmass}}
\end{center}
\end{figure}

The signal yield is determined from an unbinned maximum likelihood fit to the $B$ invariant mass distribution. The signal shape is two Gaussians with different means, and is determined from Monte Carlo with the $B$ mass resolution scaled to data in the normalisation channels. The background is a first or second order polynomial. The mass distribution and yields are shown in Figure ~\ref{bmumuhmass}. For each channel a significance greater than 6$\sigma$ is observed. The measured branching ratios are $\mathcal{B}(B^+ \to K^+ \mu^+ \mu^-) = [0.38 \pm 0.05(\mathrm{stat}) \pm 0.03(\mathrm{syst})]\times 10^{-6}$,  $\mathcal{B}(B^0 \to K* \mu^+ \mu^-) = [1.06 \pm 0.14(\mathrm{stat}) \pm 0.09(\mathrm{syst})]\times 10^{-6}$,  $\mathcal{B}(B_s \to \phi \mu^+ \mu^-) = [1.44 \pm 0.33(\mathrm{stat}) \pm 0.46(\mathrm{syst})]\times 10^{-6}$.~\cite{bmumuh} These numbers are consistent with previous results and other B-factory measurements. In the case of $B_s \to \phi \mu \mu$, this is the first observation of this decay channel and is also the rarest $B_s$ decay observed so far. 

\begin{figure}
\begin{center}
\psfig{figure=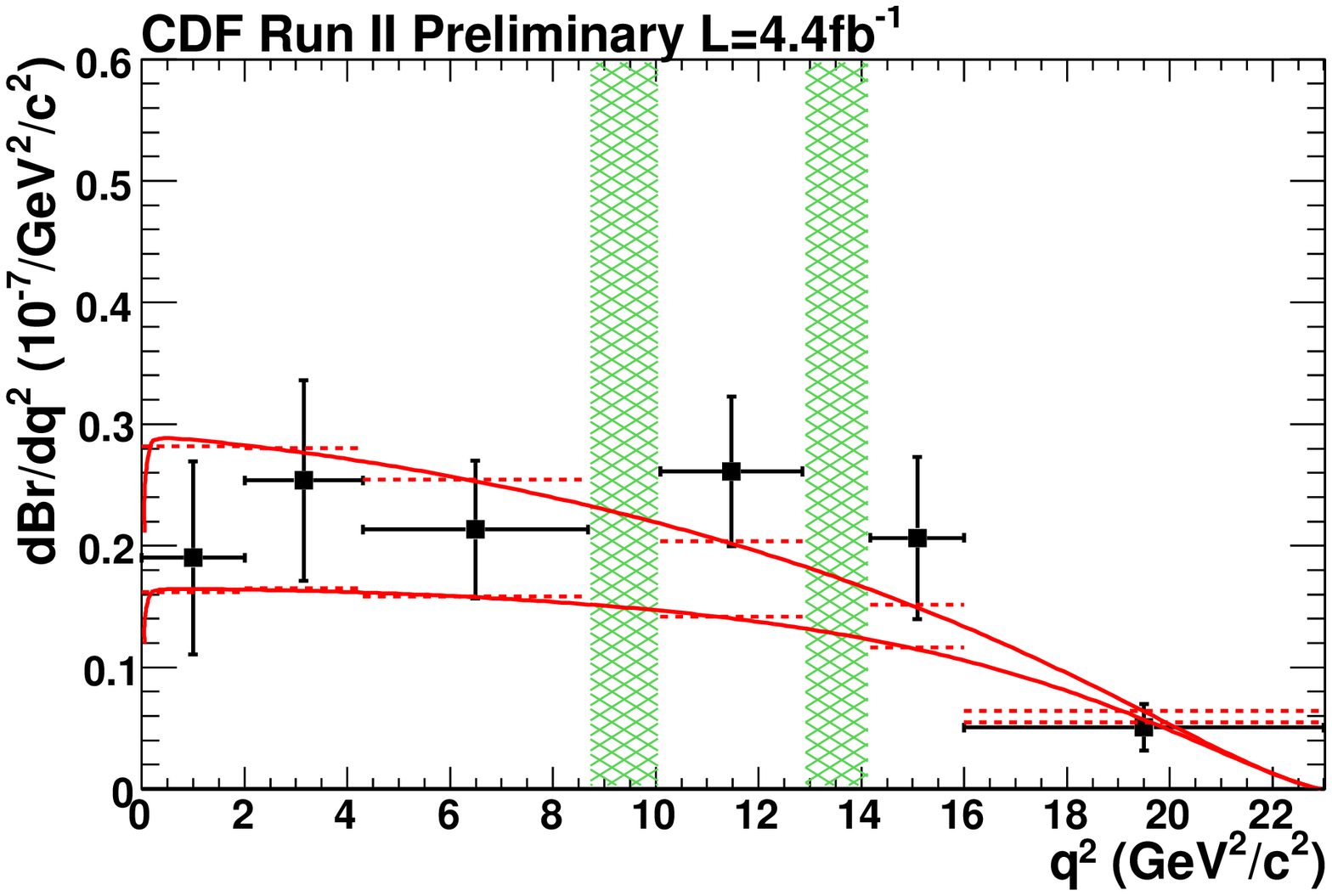,height=1.8in}
\psfig{figure=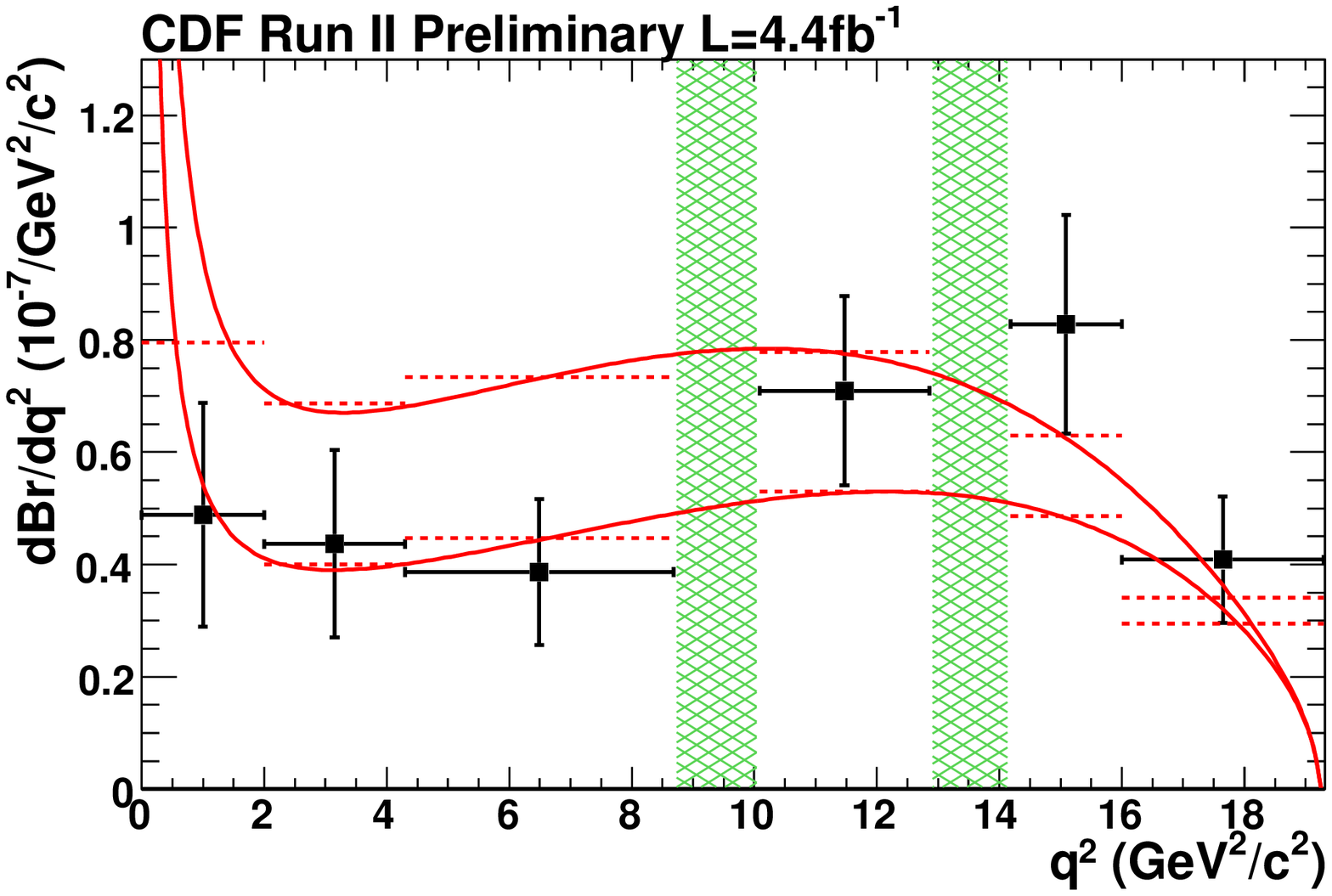,height=1.8in}
\caption{The differential branching ratios for $B^+$ (right) and $B^0$(left). The hatched regions are charmonium veto regions. Solid lines are the SM expectation.
\label{bmumudiffbr}}
\end{center}
\end{figure}

The differential branching ratio as a function of the di-muon invariant mass is determined by binning the data into 6 bins and repeating the branching ratio measurement. The 6 bins are chosen to correspond to the analysis performed at Belle to aid comparison between the two experiments. The results are shown in Fig.~\ref{bmumudiffbr}. The region consistent with the standard model is between the two lines; this band exists due to uncertainty in form factors. No inconsistency with the standard model is observed.

\begin{figure}
\begin{center}
\psfig{figure=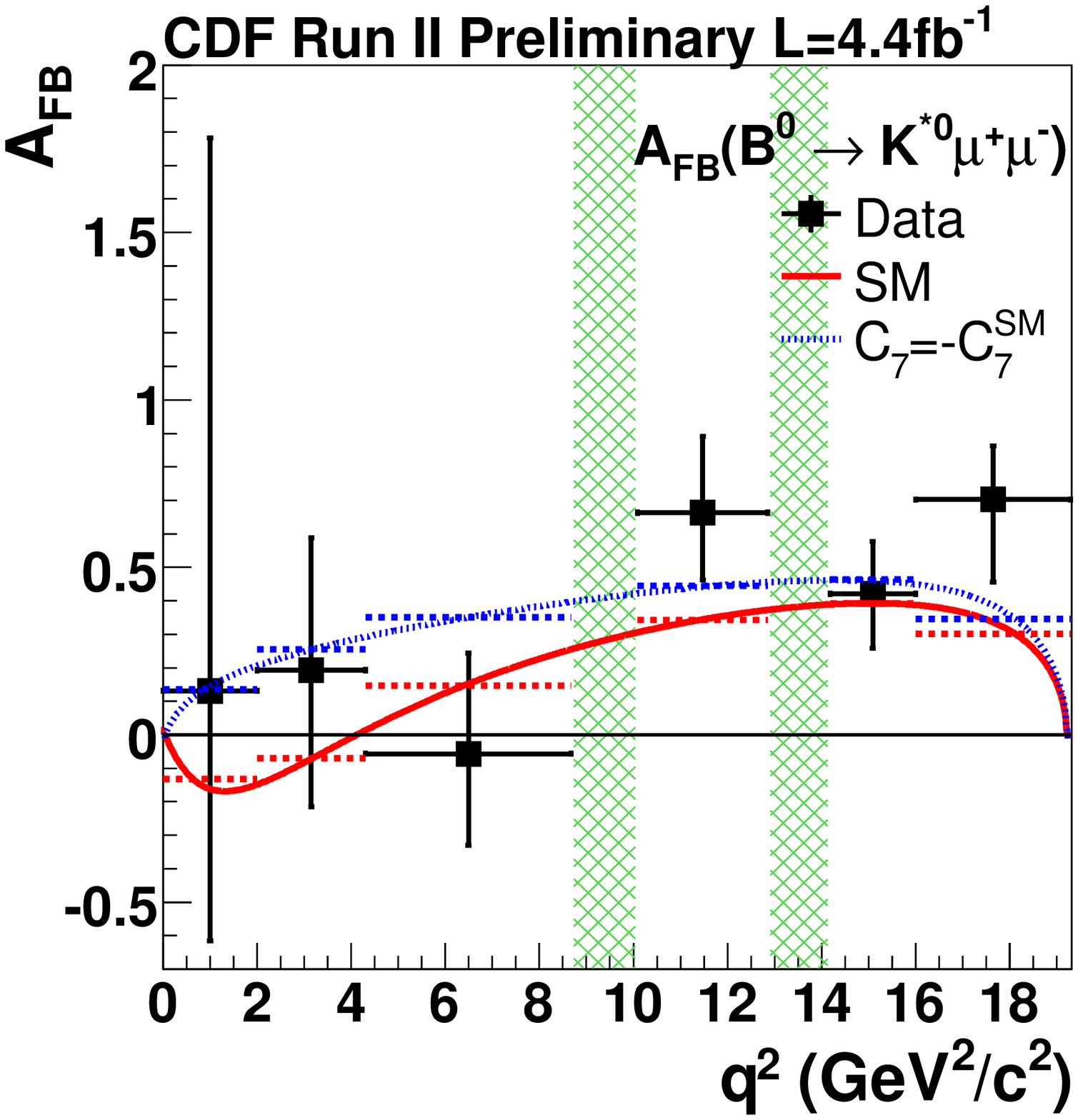,height=1.8in}
\psfig{figure=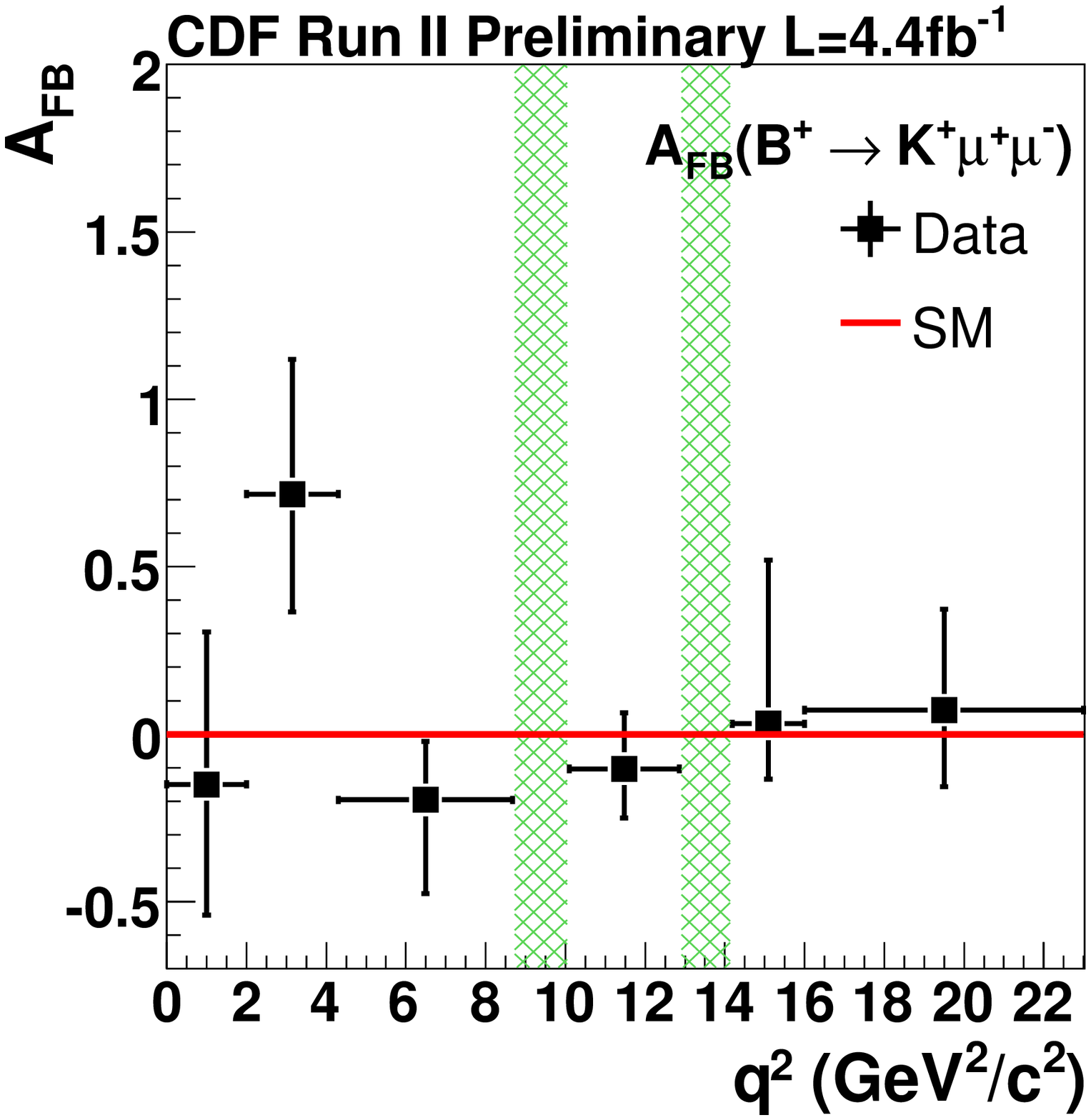,height=1.8in}
\psfig{figure=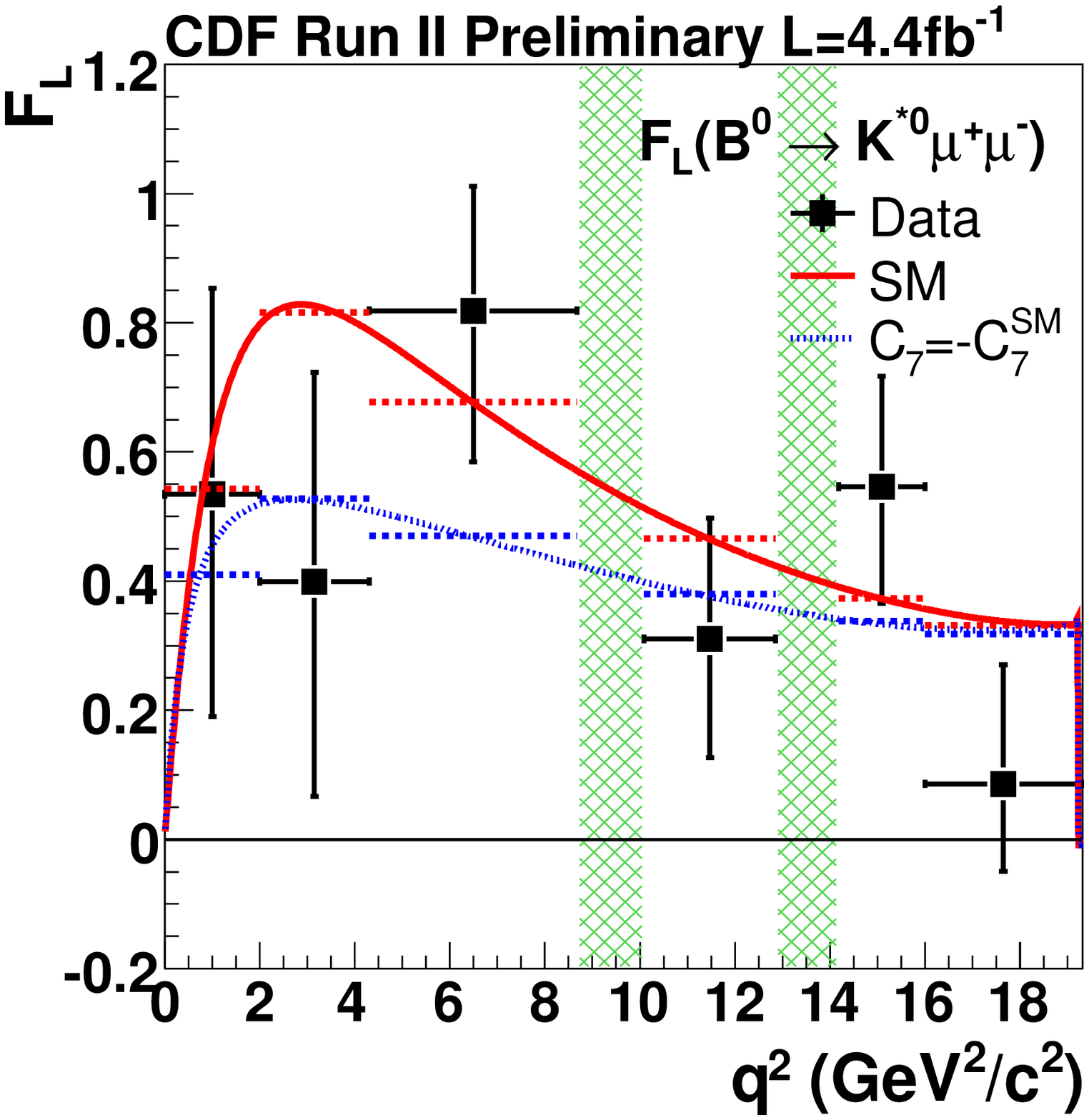,height=1.8in}
\caption{Results for $A_{FB}$ and $K^*$ polarisation. The solid (dotted) line is the standard model (a new physics model) expectation.
\label{bfab}}
\end{center}
\end{figure}

The $B^+$ and $B^0$ decays are used to determine $A_{FB}$ and the longitudinal polarisation of the $K^*$. The theoretical uncertainty on the standard model prediction has very small uncertainty and hence provides an interesting opportunity to observe the effects of new physics. $A_{FB}$ is determined from the distribution of $\cos(\theta_{\mu})$, where $\theta_{\mu}$ is the helicity angle between the $\mu^+ (\mu^{-})$ direction and the opposite of the $B(\bar{B})$ direction in the di-muon rest frame. The polarisation is measured from $\theta_K$ where this is the angle between the kaon direction and the direction opposite to the $B$ meson in the $K*$ rest frame. The fit results are shown in Fig.~\ref{bfab}, where one possible beyond the standard model scenario is also shown. The results are consistent with the standard model prediction\cite{smth} and are consistent and of similar precision to those observed at the $B$ factories.
\section{Conclusion}
The Tevatron continues to make significant contributions to the field of flavor physics. The results presented here have shown the possibilities of hadron colliders for measurements in flavor physics despite the high combinatoric backgrounds. The contribution to lifetimes will be used to better understand the interaction of quarks inside hadrons. The indirect searches of new physics through flavor physics observables remain complementary to direct searches. 

The Tevatron accelerator continues to provide a high luminosity and these analyses can be expected to be updated with 10 fb$^{-1}$. This will allow for significantly improved measurements for the measurements that are currently limited by statistical uncertainty. This will be particularly interesting for $B \to \mu \mu$ and $B \to \mu \mu h$ where hints of new physics could be observed. 

\section*{Acknowledgements}
I thank the University of Pittsburgh for the financial support that enabled my participation at this conference.

\end{document}